\documentclass{llncs}
\usepackage{graphicx}
\usepackage{hyperref} 
\usepackage[sort, numbers]{natbib}
\setcitestyle{square}
\usepackage[normalem]{ulem}
\useunder{\uline}{\ul}{}
\renewcommand\bibname{\textsc{References}}
\usepackage{url}

\usepackage{breakurl}
\usepackage{etoolbox}
\usepackage{fancyhdr} 
\let\llncssubparagraph\subparagraph
\let\subparagraph\paragraph
\usepackage[compact]{titlesec}
\let\subparagraph\llncssubparagraph

\begin{document}

\title{Atomic Loans: Cryptocurrency Debt Instruments}

\author{By: Matthew Black \inst{1} \and
TingWei Liu \inst{2} \and \\
Contributions from: Tony Cai\inst{3}}
\authorrunning{M. Black et al.}

\institute{\email{matthewjablack@gmail.com} \and
\email{tingwei890@gmail.com}\\ \and
\email{tcai534@gmail.com}}
\maketitle              
\begin{abstract}
Atomic swaps enable the transfer of value between the cryptocurrencies of various blockchains without the need to trust an intermediary. In this paper, we propose the concept of atomic loans, which utilize atomic swap technology to allow market participants to create overcollateralized debt instruments in a trustless and disintermediated manner. The primary use cases for atomic loans include enabling fiat/stablecoin access for cryptocurrency holders to participate within legacy systems, as well as enabling leverage for margin trading. We also introduce a bidding process for liquidating collateral in the case of default which provides the ability for a more fair distribution of collateral.

\end{abstract}
\section{Disintermediated Debt Instruments}

The ability to transfer value securely between the participants of a blockchain network~\cite{ref_article1} is dependent upon the cryptographic proofs and consensus protocols that are used to secure the network~\cite{ref_article2}. Yet in the absence of robust blockchain interoperability solutions in this multi-ledger world, blockchains operate in silos with users frequently depending on centralized off-chain solutions to facilitate the exchange of value across blockchains~\cite{ref_article3}.

For the purpose of lending, many centralized solutions~\cite{ref_article4} have come to fruition to allow for the creation of cryptocurrency-backed loans, allowing users to deposit  a cryptocurrency into a centralized escrow, in exchange for a fiat loan. These lending services are centralized entities with their own self-interests, and thus have the ability to seize or steal funds. In addition, there is no way to keep these entities accountable~\cite{ref_article5}, and due to the probabilistic finality of cryptocurrencies, there is no way to recover these funds~\cite{ref_article6}.

There exists alternative solutions that enable more decentralized cryptocurrency-backed loans. However, these decentralized services are still intermediated, either by smart contract owners or governance systems built into the protocol~\cite{ref_article7}. Evidently, there is a lack of solutions that enable cryptocurrency-backed loans without delegation of trust to an intermediary. 

Additionally, very few of these solutions enable loans to occur across borders and jurisdictions at the protocol level, creating silos of liquidity only accessible by a fraction of the potential users that would benefit from this type of debt instrument. This is a major threat to open, neutral, borderless, censorship-resistant networks. In this paper we introduce how atomic swaps~\cite{ref_article8} can be extended for the creation of the cross-chain debt contracts, which we have named atomic loans.  

\section{Setup}

Throughout the paper we define the two primary parties as Alice (Party A/Debtor/ Borrower) and Bob (Party B/Creditor/Lender), in addition to Charlie (Party C/Bidder), an optional third party in the case of default. For the purposes of discovery and matching, we assume that Alice and Bob have some type of communication with each other before entering into a agreement.

In this paper, we review how atomic swaps work, then show how atomic loans can be created as an extension of this underlying technology. We show various features, such as early cancellation and liquidation of collateral in the event of a default. Finally, we discuss some limitations of trustless atomic loans. 

\section{Atomic Swap}

Suppose Alice wishes to exchange ACoin for Bob’s BCoin. If Alice were to simply send a transaction with ACoin to Bob, Bob could decide to not reciprocate and keep both Alice's ACoin and his own BCoin. These methods require one party to delegate trust to the other in order to exchange value. To facilitate trustless exchange, Alice locks her ACoin's into a Hashed Time Lock Contract~\cite{ref_article9} (HTLC). In layman's terms, a HTLC enables a user to lock funds and have another party redeem them using a secret. If they do not redeem the funds within the lock time, the original party can refund the amount. In this case, Alice locks her ACoin funds for lock time $T$ with Secret $A$. Subsequently, Bob locks his BCoin funds into a HTLC for lock time $T / 2$. Alice can then redeem Bob’s BCoins in time $T / 2$. When she does so, she reveals Secret $A$, enabling Bob to redeem ACoins funds with adequate time $T$. If Alice chooses not to redeem and allows the contact to expire, then Bob can refund his BCoins after time $T / 2$, and Alice can refund her ACoins after time $T$. This process is known as an atomic swap~\cite{ref_article8}.

\section{Atomic Loan}

Notice how the process of presenting secrets in atomic swaps enables for the parties to exchange values at specific times without the need to act honestly. This framework allows us to extend the concept of atomic swaps to enable debt and repayment between parties at different intervals in a loan process, by strategically revealing secrets to represent various actions in the loan process. 

\subsection{Secrets}

Both parties generate two secrets before engaging in an atomic loan. We will call this Secret $A_1$, $A_2$ for Alice, and Secret $B_1$, $B_2$ for Bob. Additionally, in the case of a default, Charlie also generates a secret. We will call this Secret $C$. 

By having Bob reveal $B_1$ to signify acceptance of locked collateral and Alice reveal $A_1$ in the process of accepting the loan principal, the contract can determine when a participant should or should not be able to redeem funds from the contracts with the use of a HTLC using Secret $A_1$ or $B_1$. 

For example, to determine whether Bob is able to seize a predetermined percentage of the collateral (i.e. Seizable Collateral as per section~\ref{ssec:4:3}) put forth by Alice in the case of a default, we can use the following process. If Alice has revealed $A_1$ in the process of accepting the loan principal, and no other actions have occurred in the loan process, then we can assume that Alice has defaulted on the loan, and Bob has the right to claim the Seizable Collateral.

Additionally, Secrets $A_2$ and $B_2$ can be used to represent the repayment of the loan and refunding of collateral, or for the agreement of parties to liquidate the collateral in the case of default. 

Secret $C$ is also used in the case of default. During the default period, the collateral acts as a multisignature wallet~\cite{ref_article10} enabling Alice and Bob to both sign off on who the funds should be sent to. They both generate signatures for this, and once Charlie is satisfied that the signatures from Alice and Bob are valid for a multisignature transaction sending the funds to him, he reveals Secret $C$. 

\begin{table}[]
\centering
\begin{tabular}{c|l|l}
\multicolumn{1}{c|}{\textbf{Secret}} & \multicolumn{1}{c|}{\textbf{Revealed}} & \multicolumn{1}{c}{\textbf{Used}} \\ \hline
$A_1$ & \begin{tabular}[c]{@{}l@{}}when Alice accepts and withdraws \\ loan\end{tabular}                                                                                                                                                                                                                                                         & \begin{tabular}[c]{@{}l@{}}by Bob to seize the collateral in the \\ case that Alice defaults on the loan \\ repayment\end{tabular}                                                                         \\ \hline
$A_2$     & \begin{tabular}[c]{@{}l@{}}in bidding period if Alice has \\ already signed multisignature \\ transaction to accept \\ Charlie’s bid\end{tabular}                                                                                                                                                                                                          & \begin{tabular}[c]{@{}l@{}}by Charlie, along with Secret $B_2$, Alice \\  signature for multisignature transaction \\  and Bob signature for multisignature \\ transaction to redeem Alice’s collateral\\ in the bidding period\end{tabular} \\ \hline
$B_1$     & when Bob accepts Alice’s locked collateral                                                                                                                                                                                                                                                                                               & \begin{tabular}[c]{@{}l@{}}by Alice to withdraw the loan \\ principal after Bob has accepted her \\ locked collateral\end{tabular}                                                                         \\ \hline
$B_2$     & \begin{tabular}[c]{@{}l@{}}if Bob allows HTLC containing \\ loan principal to expire and refunds principal\\ {\ul OR} if Bob redeems loan repayment \\ HTLC containing principal plus interest\\ {\ul OR} in bidding period, if Bob has already \\ signed the multisignature transaction \\ for Alice's collateral and accepts bid of Charlie \end{tabular} & \begin{tabular}[c]{@{}l@{}}by Charlie along with Secret $A_2$, \\ Alice signature for multisignature \\ transaction, and Bob signature for \\ multisignature transaction to redeem \\ Alice’s collateral in the bidding period\end{tabular} \\ \hline
$C$      & \begin{tabular}[c]{@{}l@{}}when Charlie has verified that both \\ signatures provided by Alice and \\ Bob in the bidding period are valid\end{tabular}                                                                                                                                                                                   & \begin{tabular}[c]{@{}l@{}}by Alice or Bob to allow them to \\ redeem the bid locked by Charlie\end{tabular}                                                                                              
\end{tabular}
\end{table}

\subsection{Loan Process}
For ease of reading, we have broken the various parts of the atomic loan process into four periods, loan, bidding, seizure, and refund periods. These serve to enable easier understanding of the loan process and split up the outcomes that could arise from an atomic loan. 
\subsubsection{Loan Period}

The loan period is designated to handle the loan withdrawal and repayment process. If the repayment is successful, the other periods are not necessary. Alice first requests a loan, and sends the loan information to Bob. Bob then locks the loan amount into a HTLC. Once Alice is satisfied that the principal of the loan has been locked, she will lock their collateral into a HTLC. Once Bob is satisfied that the collateral has been locked, he releases the funds to Alice. Alice then takes the funds for the duration of the loan period, and then pays back the principal plus interest at the end of the loan period. Once Bob is satisfied that the loan has been repaid, he reclaims the principal plus interest and enables Alice to refund her collateral.  

\subsubsection{Bidding Period}
In the case that Alice does not repay the loan, or Bob suspects that the loan collateral has gone below the agreed upon minimum collateralization ratio, Alice and Bob have the option to enter into a `Bidding Period' which liquidates the collateral, and repays Bob for the principal, interest plus a liquidation fee. 

\subsubsection{Seizure Period}
If Alice or Bob decide not to opt into liquidation of the collateral, then Bob will have the option to claim the Seizable Collateral. This is defined before the loan begins. This, of course, opens Bob up to market fluctuation. 

\subsubsection{Refund Period}
If Bob stops responding during the previous periods after Alice has locked their collateral, then Alice can still recover their collateral during this period as a last resort. 

\subsection{Collateral Locking} \label{ssec:4:3}

When Alice and Bob agree to the terms of the loan, they also agree to the percentage of collateral that Bob will be entitled to in the event that Alice or Bob decide not to opt into liquidation of the collateral in the Bidding Period. Since inputs and outputs cannot be predetermined in a typical UTXO~\cite{ref_article11} scheme, we lock these two values separately into distinct P2SH contracts, into the Seizable Collateral, that Bob has the ability to seize in the Seizure Period, and Refundable Collateral, that Alice has the ability to refund in the Seizure Period. Both the Seizable and Refundable can be claimed by Charlie following the conclusion of the Bidding Period (see section~\ref{ssec:5:5}).

\section{Protocol Specification}

\subsection{Becoming a Lender}
In order for Bob to become a Lender in an atomic loan, he must first generate two secrets $B_1$, $B_2$, and then lock the principal of the loan into a HTLC. 

\subsection{Becoming a Borrower}
Alice can become a Borrower in an atomic loan by ensuring they have the necessary collateral in ACoin to cover a loan in BCoin with a collateral-to-debt ratio that both parties have agreed upon. This can be based on protocols with similar collateral-to-debt ratio requirements~\cite{ref_article12} such as a 150\% collateral-to-debt ratio. In addition, the reputation of an individual based on their performance within past atomic loans can help determine this ratio. This reputation can be based on the address alone, which retains the pseudo-anonymous nature of the loan. Alice then generates two secrets, Secret $A_1$ and Secret $A_2$. Alice then locks their collateral into two HTLC's.

\subsection{Initiating Loan}
Once Bob is satisfied that Alice has locked her collateral into an HTLC, Bob can provide the Secret $B_1$ to Alice. Alice can then use Secret $B_1$ to redeem the principal of the loan from the atomic loan contract, in the process revealing Secret $A_1$, which acts as proof that she withdrew the loan principal.  

\subsection{Repaying Loan}
\subsubsection{Successful Repayment}
Once Alice’s loan term has finished, Alice repays the principal plus interest agreed upon and locks it into the loan contract. In order for Bob to redeem, he must reveal Secret $B_2$. At this point, Alice has both Secret $B_1$ and Secret $B_2$. Alice now has all the elements required to refund her initial collateral. 

\subsubsection{Unsuccessful repayment}
If Alice does not repay back the loan at the end of the loan period, either Alice or Bob can initiate the bidding process. 
\subsubsection{Successful Repayment Without Reciprocation}
In the case that Alice's loan term finishes and she repays the loan, and Bob decides not to accept the loan and release Secret $B_2$, there is no way for Alice to refund her collateral. To ensure that Bob does not simply start the bidding process and profit from the additional liquidation fees accrued, the contract can detect if Alice has paid back the loan to stop the bidding process from occurring. The longer Bob waits to reciprocate, the less interest he gets from the loan. At a certain point, the lock time will run out on the contract, and Bob must resort to redeeming the Seizable Collateral in the Seizure Period to compensate himself for the loan capital he provided. The timeout period until Bob can collect the Seizable Collateral should be sufficiently long to ensure that present value of money is a deterrent from choosing this last resort route.

\subsection{Bidding Process in the Case of Default} \label{ssec:5:5}
In the event that Alice does not pay back the loan, the bidding process serves as a fair way to liquidate the collateral in order to ensure that Bob gets his fair share of the collateral, and that Alice gets access to the correct valuation of the remaining collateral. 

This process only occurs in the case that Alice does not pay back the loan. Either Alice or Bob can initiate the bidding process. This is done by calling a function in the contract, changing the state of the contract to bidding, which will allow any third party, in addition to Alice or Bob, to partake in the bidding process (since even if Alice or Bob aren't allowed in the bidding process, they could just use another address anyway). 

In this process, participants will stake a BCoin amount higher than the previous bidder, which then triggers a refund of the BCoin amount staked as part of the previous highest bid to its respective bidder. When bidding is closed, the highest bidder’s (Charlie's) funds will be locked into an HTLC to give adequate time for Alice and Bob to sign a multisignature transaction to allow for Charlie to redeem both the Refundable and Seizable Collateral. 

\subsection{Charlie's Acquisition of Alice's Collateral}
For the purposes of allowing Charlie to obtain the ACoin collateral through a swap with his staked BCoin, Alice and Bob must both sign the multisignature transaction to send funds to Charlie's ACoin blockchain address. This is for the purpose of preventing front-running. Once Charlie is satisfied that the multisignature transaction signatures are valid, he can reveal a Secret $C$. After this, Alice and Bob must also reveal Secret $A_2$ and $B_2$.

Alternatively, for a higher cost, the contract itself on the BCoin blockchain could verify the signatures, rather than requiring Secret $C$, provided that the original ACoin transaction was represented in some manner in the BCoin blockchain. However, doing signature validation of a multisignature from another blockchain would be quite costly, and would only be worthwhile for large transactions (provided there was a way to do it that did not exceed any transaction cost restraints). 

Once the contract verifies that Secret $A_2$ and $B_2$ have been revealed, the funds that were staked by Charlie will be redeemable for Alice and Bob, with Bob receiving principal, interest and an additional fee, and Alice receiving the remainder. Charlie will be able to use the multisignature transaction signatures as well as Secret $A_2$ and $B_2$ to redeem the Refundable and Seizable Collateral.

\subsubsection{Collateral Acquisition where Alice or Bob is Charlie} 

We describe the scenario where either Alice or Bob is Charlie in this swap. After both Alice and Bob sign the multisignature transaction to send funds to Charlie, the “double-agent” party might decide not to reveal their secret ($A_2$ or $B_2$). In this case, they will end up revealing the secret anyway on the ACoin blockchain if they decide to try and redeem the collateral (both Refundable and Seizable) as Charlie. The other party who did reveal their secret can then call a function on the BCoin blockchain revealing the unrevealed secret in the contract so that the double-agent (non-secret revealer) does not refund the bid.  

For the purpose of this example, let’s assume Alice is secretly Charlie. In the case that both parties provide a valid signature for the multisignature transaction with the output being Charlie's ACoin address, and Bob reveals Secret $B_2$. Alice can redeem the collateral as Charlie, by using the signature of herself and Bob, as well as Secret $B_2$ that was revealed by Bob, and Secret $A_2$ which only she has access to. However, once the collateral is redeemed, Bob can scan the ACoin chain and see that Secret $A_2$ has been revealed, and subsequently reveal it on the BCoin chain by calling a specific function dedicated to this scenario, in order for him to be able to redeem Charlie's bid.

\subsection{Collateral Seizure in the Case of Default and Failed Bidding}

In the case where Alice or Bob decide not to sign or reveal either of their secrets at the conclusion of bidding, Bob will have the opportunity to redeem the Seizable Collateral in the Seizure Period. This is possible after a set amount of time allocated to the Bidding Period (preferably longer, to give incentive for both parties to opt for bidding process). The amount of funds allocated to Refundable versus Seizable Collateral is decided before entering the loan contract. There is a lot of consideration around what percentage of the total collateral should be allocated to Seizable Collateral and be redeemable by Bob since market fluctuation could give incentive for either party to opt for Collateral Seizure over the bidding process (see diagram below). There is also the current market to take into account (i.e. bull market vs bear market). This will need to be decided by both parties before entering the loan contract. During the Seizure Period, Bob will be able to redeem the collateral, using Secret $A_1$. Alice is subsequently able to redeem the Refundable Collateral during the Seizure Period as well.

\subsection{Refunding Collateral as a Last Resort}

In the case that Alice locks her collateral, and is unable to take out the loan amount because Bob never reveals any secrets (potentially because he is unhappy with the amount of collateral Alice locked in), Alice will have the opportunity to refund her Seizable and Refundable  Collateral in the Refund Period. This is a last resort in the case that Bob halts interaction with the blockchain altogether. 

\hspace*{\fill} \\

\begin{table}[]
\centering
\def\arraystretch{1.5}%
\setlength\tabcolsep{10pt}
\begin{tabular}{l|l}
\multicolumn{1}{c|}{\textbf{Repayment Outcome}}                                                                  & \multicolumn{1}{c}{\textbf{Alice}}                                                                                             \\ \hline
\begin{tabular}[c]{@{}l@{}}Payment of loan \\ principal plus interest\end{tabular} & \begin{tabular}[c]{@{}l@{}}ACoin value increases \\ Time value of money\\ Reputation\end{tabular} \\ \hline
Default on loan                                                                    & ACoin value decreases                                                                            
\end{tabular}
\end{table}

\begin{center}
Incentives for Alice to payback the loan principal plus interest
\end{center}

\begin{table}[]
\centering
\def\arraystretch{1.5}%
\setlength\tabcolsep{10pt}
\begin{tabular}{l|l}
\multicolumn{1}{c|}{\textbf{Repayment Outcome}}                                                                  & \multicolumn{1}{c}{\textbf{Bob}}                                                                                               \\ \hline
\begin{tabular}[c]{@{}l@{}}Accept payment of loan \\ principal plus interest\end{tabular} & \begin{tabular}[c]{@{}l@{}}ACoin value decreases \\ Time value of money\\ Reputation\end{tabular} \\ \hline
Wait for seizure period                                                                   & ACoin value increases                                                                            
\end{tabular}
\end{table}

\begin{center}
Incentives for Bob to accept the loan repayment from Alice
\end{center}

\begin{table}[]
\centering
\def\arraystretch{1.5}%
\setlength\tabcolsep{10pt}
\begin{tabular}{l|l|l}
\multicolumn{1}{c|}{\textbf{Default Outcome}}   & \multicolumn{1}{c|}{\textbf{Alice}}                                                                                      & \multicolumn{1}{c}{\textbf{Bob}}                                                                                        \\ \hline
Enter bidding period    & \begin{tabular}[c]{@{}l@{}}ACoin value increases\\ Time value of money\\ Reputation\end{tabular} & \begin{tabular}[c]{@{}l@{}}ACoin value decreases\\ Time value of money\\ Reputation\end{tabular} \\ \hline
Wait for seizure period & ACoin value decreases                                                                            & ACoin value increases                                                                           
\end{tabular}
\end{table}

\begin{center}
Incentives for each party to choose an outcome in the case that Alice defaults on the loan
\end{center}

\noindent
{\ul Legend:}

\noindent
{\ul ACoin value increases/decreases:} value changes relative to BCoin value

\noindent
{\ul Time value of money:} Coin is worth more in the present than in the future

\noindent
{\ul Reputation:} Good actions will result in favorable interest rates and counterparties in the future

\section{Other Considerations}

Similar to any long term contracts on the blockchain, there is always the possibility that there may be a change to the underlying blockchain, such as a hard or soft fork, in which a change occurs in the underlying hashing or signing algorithm of the chains involved. There is the possibility that Alice and Bob could cooperate in order for the loan to complete properly, but there would need to be delegation of trust. In the case that a fork occurs affecting the hashing algorithm on ACoin, if Alice defaults, and isn't cooperative, she can exercise her ability to refund the collateral amount during the refund period. However, Bob can also harm Alice by revealing the secret for accepting repayment after a fork has occurred, affecting her ability to refund her collateral until the refund period. 

The atomic loan process is unlikely to be user friendly if dealing directly with the blockchain since it requires an understanding of atomic swaps, hashing, and basic knowledge of the underlying blockchain. 

The bidding process requires the BCoin blockchain to have smart contract capabilities in order to allow detection of Alice and Bob's acceptances of Charlie's bid which is a prerequisite for the redemption of the liquidated funds.

There is no need for Secret $B_1$ to be in the form of a secret and secret hash pair for the purposes of the atomic loan. If the BCoin blockchain has smart contract capabilities, then it is possible for this to be represented as a state update. 
In order to ensure that the swaps expire in the proper time frame, timelocks must be set in linux epoch timestamps, rather than use block height. This is because it is difficult to predict the future blocks between chains. 

There now exists various different implementations of stablecoins, or price stable cryptocurrencies that are pegged against fiat currencies. While stablecoins may serve to greatly reduce the volatility of the asset upon which the loan principal is denominated in, not all stablecoins are created equal and can take vastly different approaches in the implementation of peg. As such, participants who choose to participate in stablecoin-denominated loans must be willing to trust the underlying peg of the stablecoin.

\subsection{Similar projects}
Similar projects for decentralized loan contracts focus primarily on ERC20 tokens on the Ethereum Blockchain~\cite{ref_article13}. In our approach, we outline a protocol that can be used between different blockchains, with the requirements that one of the blockchains be smart contract compatible, and that they both support the creation of HTLCs with the same hash function. Additionally, the blockchains do not need to be Turing complete like Ethereum as long as the atomic loan does not make use of the bidding process. Others have built projects where the entirety of the loan is intermediated by the governance mechanism. Loans on such protocols require the delegation of trust to holders of the governance token as well as the governance mechanism as a whole which is not controlled by the underlying consensus mechanism of the blockchain it runs on~\cite{ref_article7}. 

\section{Conclusion}
The adoption and development of disintermediated blockchain applications is an ongoing process. While it is still not clear what features blockchains should support to enable the development of disintermediated applications on top of it, what is clear is that current solutions to obtain fiat/stablecoin liquidity via cryptocurrency loans leaves much to be desired in terms of disintermediation and the delegation of trust. We have demonstrated how the underlying concepts and technology of atomic swaps can be used for alternative purposes such as atomic loans. Atomic loans provide a sound and resilient alternative for crypto holders to leverage their holdings to enable liquidity in a disintermediated manner and freely participate within the constraints of legacy financial systems.

%
%
%
%

\begin{flushleft}
\renewcommand\bibname{References \vspace*{-10mm}}
  
\bibliographystyle{rsc}

\end{flushleft}

\section{Appendix}
\label{sec:appendix}

\subsection{Contract Diagrams} 

This section contains example diagrams for the contracts described in section 4. Note that boxes correspond to transactions where the fill color indicates which blockchain the transaction takes place on, while the border color indicates who can publish the transaction.

\begin{figure}
\includegraphics[
  width=33cm,
  height=19cm,
  keepaspectratio,
]{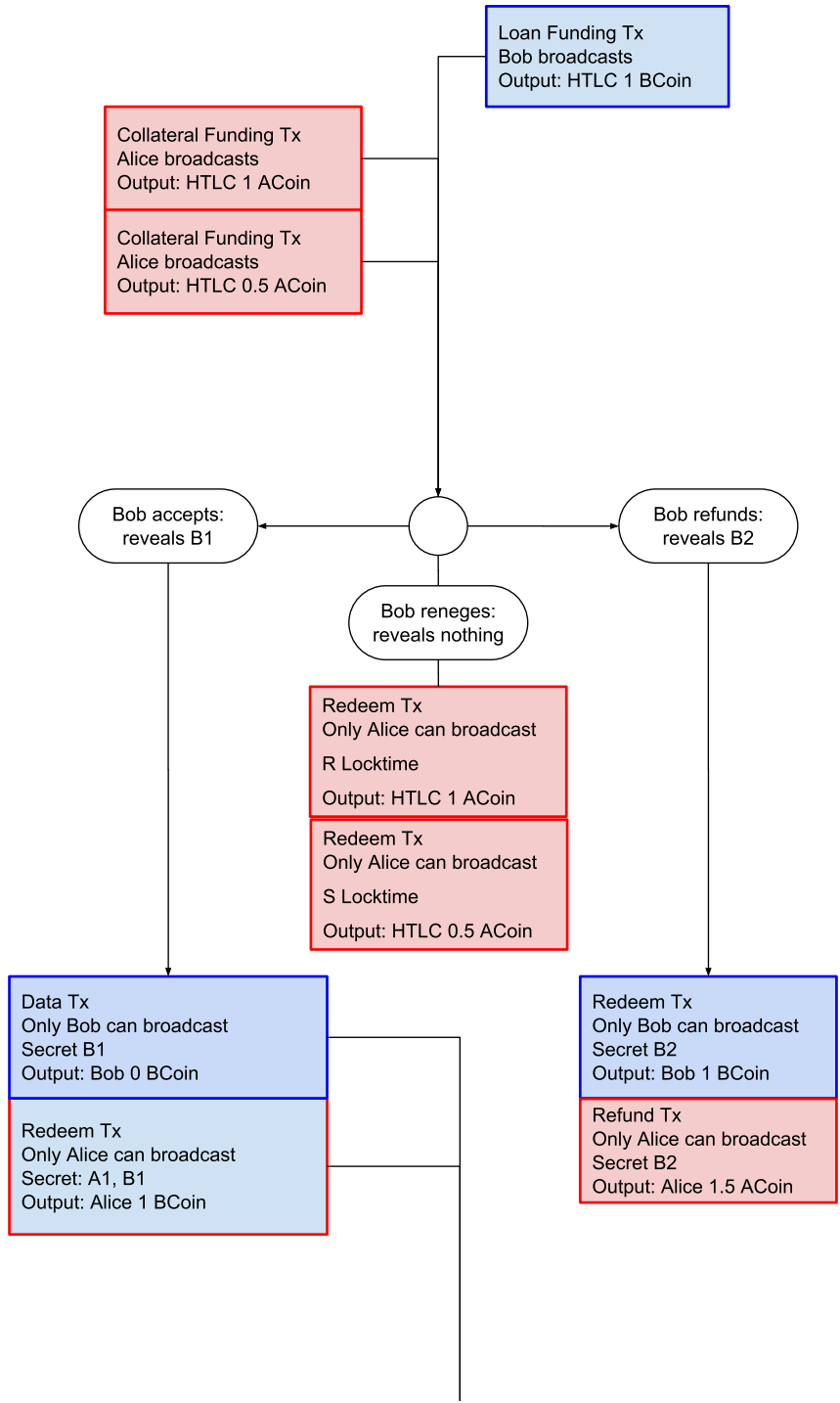}
\caption{Basic atomic loan construction.} \label{Figure 1}
\end{figure}

\begin{figure}
\centering
\includegraphics[
  width=33cm,
  height=19cm,
  keepaspectratio,
]{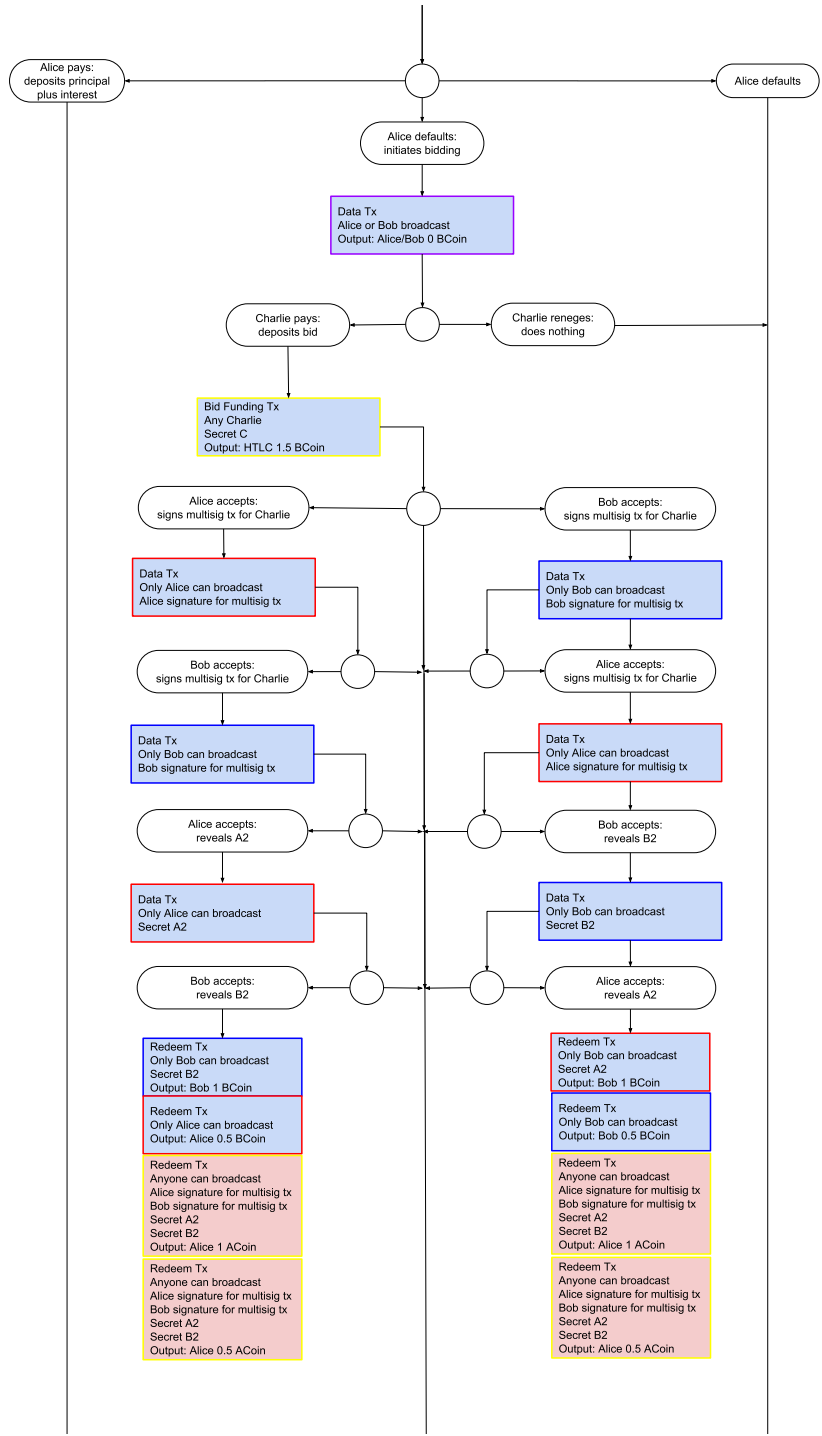}
\caption{Bidding period.} \label{Figure 1}
\end{figure}

\begin{figure}
\includegraphics[width=\textwidth]{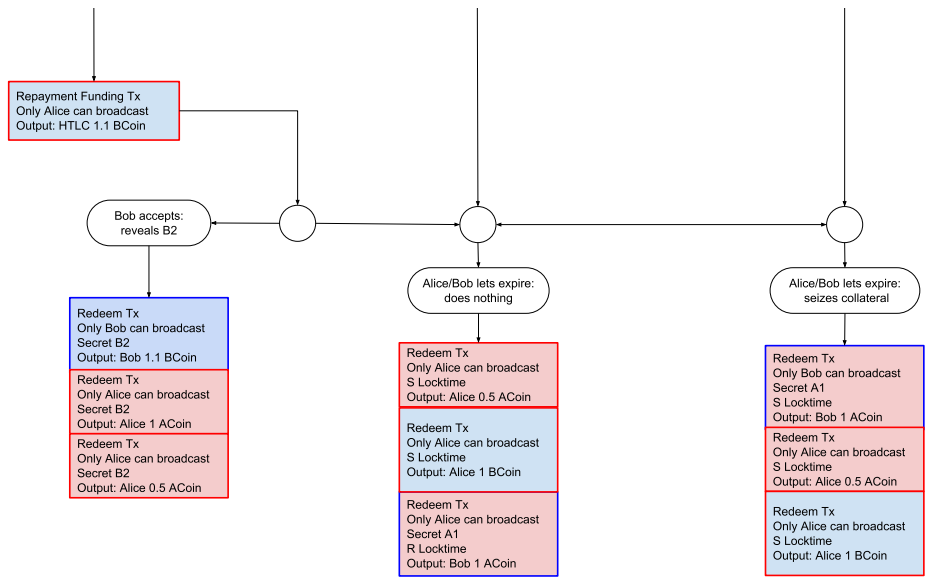}
\caption{Seizure and refund periods.} \label{Figure 1}
\end{figure}

\end{document}